\documentclass[preprint]{revtex4}

\usepackage{graphicx}

\draft

\begin{document}
\titlepage
\title{Palatini Formulation of Modified Gravity with $\ln R$ Terms}

\author{Xin-He Meng$^{1,2,3}$ \footnote{mengxh@phys.nankai.edu.cn}
 \ \ Peng Wang$^1$ \footnote{pwang234@sohu.com}
} \affiliation{1.  Department of Physics, Nankai University,
Tianjin 300071, P.R.China \\2. Institute of Theoretical Physics,
CAS, Beijing 100080, P.R.China \\3. Department of Physics,
University of Arizona, Tucson, AZ 85721}

\begin{abstract}
Recently, corrections to the standard Einstein-Hilbert action are
proposed to explain the current cosmic acceleration in stead of
introducing dark energy. We discuss the Palatini formulation of
the modified gravity with a $\ln R$ term suggested by Nojiri and
Odintsov. We show that in the Palatini formulation, the $\ln R$
gravity can drive a current exponential accelerated expansion and
it reduces to the standard Friedmann evolution for high redshift
region. We also discuss the equivalent scalar-tensor formulation
of the theory. We indicate that the $\ln R$ gravity may still have
a conflict with electron-electron scattering experiment which
stimulates us to pursue a more fundamental theory which can give
the $\ln R$ gravity as an effective theory. Finally, we discuss a
problem faced with the extension of the $\ln R$ gravity by adding
$R^m$ terms.
\end{abstract}

\maketitle

\textbf{1. Introduction}

That our universe expansion is currently in an accelerating phase
now seems well-established. The most direct evidence for this is
from the measurement of type Ia supernova \cite{Perlmutter}. Other
indirect evidences such as the observations of CMB by the WMAP
satellite \cite{Spergel}, large-scale galaxy surveys by 2dF and
SDSS also seem supporting this.

But now the mechanism responsible for this acceleration is not
very clear. Many authors introduce a mysterious cosmic fluid
called dark energy to explain this (see Ref.\cite{Peebles,
Carroll-de, Padmanabhan} for a review). On the other hand, some
authors suggest that maybe there does not exist such mysterious
dark energy, but the observed cosmic acceleration is a signal of
our first real lack of understanding of gravitational physics
\cite{Lue}. An example is the braneworld theory of Dvali et al.
\cite{Dvali}.

Recently, some authors proposed to add a $R^{-1}$ term in the
Einstein-Hilbert action to modify the General Relativity (GR)
\cite{Carroll, Capozziello}. It is interesting that such terms may
be predicted by string/M-theory \cite{Odintsov3}. It was shown in
their work that this additional term can give accelerating
solutions of the field equations without dark energy. Based on
this modified action, Vollick \cite{Vollick} used Palatini
variational principle to derive the field equations. In the
Palatini formalism, instead of varying the action only with
respect to the metric, one views the metric and connection as
independent field variables and vary the action with respect to
them independently. This would give second order field equations.
In the original Einstein-Hilbert action, this approach gives the
same field equations as the metric variation. For a more general
action, those two formalism are inequivalent, they will lead to
different field equations and thus describe different physics
\cite{Volovich}. Flanagan \cite{Flanagan} derived the equivalent
scalar-tensor theory of the Palatini formulation. Furthermore, in
Ref.\cite{Flanagan2}, Flanagan derived the equivalent
scalar-tensor theory of a more general modified gravity framework.
Those results are very important and fundamental for the Palatini
formalism. We will apply his framework in Sec.3 to discuss the
$\ln R$ gravity. In Ref.\cite{Dolgov}, Dolgov and Kawasaki argued
that the fourth order field equations following from the metric
variation suffer serious instability problem. If this is indeed
the case, the Palatini approach appears even more appealing,
because the second order field equations following from Palatini
variation are free of this sort of instability \cite{Wang1}.
However, the most convincing motivation to take the Palatini
formalism seriously is that the field equations following from it
fit the SN Ia data at an acceptable level \cite{Wang1}. An
extension of the $1/R$ theory, the $R+1/R+R^2$ theory has been
discussed in metric formation by Nojiri and Odintsov
\cite{Odintsov2}. It is shown that such an extension may explain
both the current acceleration and early inflation and it may
resolve the instability of the original $1/R$ gravity. Its
Palatini formation is discussed in Ref.\cite{Wang2}.
Interestingly, in the Palatini formation, while it can still drive
a current acceleration, adding a $R^2$ term can not drive a early
inflation. The difference of metric formation and Palatini
formation is thus quite obvious. But now we still can not tell
which one is physical.

In Ref.\cite{Odintsov}, Nojiri and Odintsov presented another
effort in this direction to modify gravity theory. They added a
$\ln R$ term to the Einstein-Hilbert action. They considered the
metric formation of this theory and concluded that such a theory
can derive an accelerated expansion. By the above considerations,
we think it is worth further investigating  of the Palatini
formulation of this $\ln R$ theory.

This paper is arranged as follows: in Sec.2 we derive the Modified
Friedmann (MF) equation in Palatini formulation of the $\ln R$
theory and discuss several of its features; in Sec.3 we discuss
the equivalent scalar-tensor formulation of the $\ln R$ theory and
an extension of the $\ln R$ theory also suggested by Nojiri and
Odintsov \cite{Odintsov}; Sec.4 is devoted to conclusions and
discussions.

\textbf{2. The model and the Modified Friedmann equation}

Firstly, we briefly review deriving field equations from a
generalized Einstein-Hilbert action by using Palatini variational
principle. See Refs. \cite{Vollick, Wang1, Wang2, Volovich} for
details. We will follow the sign conventions of
Ref.\cite{Weinberg} in this paper.

The field equations follow from the variation in Palatini approach
of the generalized Einstein-Hilbert action
\begin{equation}
S=-\frac{1}{2\kappa}\int{d^4x\sqrt{-g}L(R)}+\int{d^4x\sqrt{-g}L_M}\label{action}
\end{equation}
where $\kappa =8\pi G$, $L$ is a function of the scalar curvature
$R$ and $L_M$ is the Lagrangian density for matter.

Varying with respect to $g_{\mu\nu}$ gives
\begin{equation}
L'(R)R_{\mu\nu}-\frac{1}{2}L(R)g_{\mu\nu}=-\kappa
T_{\mu\nu}\label{2.2}
\end{equation}
where a prime denotes differentiation with respect to $R$ and
$T_{\mu\nu}$ is the energy-momentum tensor given by
\begin{equation}
T_{\mu\nu}=-\frac{2}{\sqrt{-g}}\frac{\delta S_M}{\delta
g^{\mu\nu}}\label{2.3}
\end{equation}
We assume the universe contains dust and radiation, denoting their
energy densities as $\rho_m$ and $\rho_r$ respectively, thus
$T^{\mu}_{\nu}=\{-\rho_m-\rho_r,p_r,p_r,p_r\}$ and
$T=g^{\mu\nu}T_{\mu\nu}=-\rho_m$ because of the relation
$p_r=\rho_r/3$.

In the Palatini formulation, the connection is not associated with
$g_{\mu\nu}$, but with $h_{\mu\nu}\equiv L'(R)g_{\mu\nu}$, which
is known from varying the action with respect to $\Gamma
^{\lambda}_{\mu\nu}$. Thus the Christoffel symbol with respect to
$h_{\mu\nu}$ is given by
\begin{equation}
\Gamma
^{\lambda}_{\mu\nu}=\{^{\lambda}_{\mu\nu}\}_g+\frac{1}{2L'}[2\delta
^{\lambda}_{(\mu}\partial
_{\nu)}L'-g_{\mu\nu}g^{\lambda\sigma}\partial
_{\sigma}L']\label{Christoffel}
\end{equation}
where the subscript $g$ signifies that this is the Christoffel
symbol with respect to the metric $g_{\mu\nu}$.

The Ricci curvature tensor is given by
\begin{eqnarray}
R_{\mu\nu}=R_{\mu\nu}(g)-\frac{3}{2}(L')^{-2}\nabla _{\mu}L'\nabla
_{\nu}L' +(L')^{-1}\nabla _{\mu}\nabla
_{\nu}L'+\frac{1}{2}(L')^{-1}g_{\mu\nu}\nabla _{\sigma}\nabla
^{\sigma}L'\label{Ricci}
\end{eqnarray}
and
\begin{equation}
R=R(g)+3(L')^{-1}\nabla _{\mu}\nabla ^{\mu}
L'-\frac{3}{2}(L')^{-2}\nabla_{\mu}L'\nabla^{\mu}L'\label{scalar}
\end{equation}
where $R_{\mu\nu}(g)$ is the Ricci tensor with respect to
$g_{\mu\nu}$ and $R=g^{\mu\nu}R_{\mu\nu}$. Note by contracting
(\ref{2.2}), we get:
\begin{equation}
L'(R)R-2L(R)=-\kappa T\label{R(T)}
\end{equation}
Assume we can solve $R$ as a function of $T$ from (\ref{R(T)}).
Thus (\ref{Ricci}), (\ref{scalar}) do define the Ricci tensor with
respect to $h_{\mu\nu}$.

Then we review the general framework of deriving modified
Friedmann equation in Palatini formulation \cite{Wang1}. Consider
the Robertson-Walker metric describing the cosmological evolution,
\begin{equation}
ds^2=-dt^2+a(t)^2(dx^2+dy^2+dz^2)\label{metric}
\end{equation}
We only consider a flat metric, which is favored by present
observations \cite{Spergel}.

From Eqs.(\ref{metric}) and (\ref{Ricci}), we can get the
non-vanishing components of the Ricci tensor:
\begin{equation}
R_{00}=3\frac{\ddot{a}}{a}-\frac{3}{2}(L')^{-2}(\partial_0{L'})^2+\frac{3}{2}(L')^{-1}\nabla_0\nabla_0L'\label{R00}
\end{equation}
\begin{eqnarray}
R_{ij}=-[a\ddot{a}+2\dot{a}^2+(L')^{-1}\Gamma^0_{ij}\partial_0L'
+\frac{a^2}{2}(L')^{-1}\nabla_0\nabla_0L']\delta_{ij}\label{ij}
\end{eqnarray}

Substituting equations (\ref{R00}) and (\ref{ij}) into the field
equations (\ref{2.2}), we can get
\begin{equation}
6H^2+3H(L')^{-1}\partial_0L'+\frac{3}{2}(L')^{-2}(\partial_0L')^2=\frac{\kappa
(\rho+3p)-L}{L'}\label{aa}
\end{equation}
where $H\equiv \dot{a}/a$ is the Hubble parameter, $\rho$ and $p$
is the total energy density and total pressure respectively..
Assume we can solve $R$ in term of $T$ from Eq.(\ref{R(T)}),
substituting it to the expression for $L'$ and $\partial_0L'$, we
can get the MF equation.

Now we turn to the consideration of the following modified
Einstein-Hilbert action suggested by Nojiri and Odintsov
\cite{Odintsov}
\begin{equation}
L(R)=R-\beta\ln \frac{R}{-\alpha}\label{lagrangian}
\end{equation}
Since our interest is to explain cosmic acceleration, we will
assume $R<0$ in this paper, i.e. de Sitter space. Thus $\alpha>0$.

The contracted field equation (\ref{R(T)}) now reads:
\begin{equation}
f(R)\equiv \frac{R}{-\beta}+2\ln\frac{R}{-\alpha}-1=-\kappa
T/\beta=\frac{\kappa\rho_m}{\beta}\label{contract}
\end{equation}
If $\beta>0$, $f(R)$ is a monotonically decreasing function and we
have $\lim_{R\rightarrow 0}f(R)\rightarrow -\infty$ and
$\lim_{R\rightarrow -\infty}f(R)\rightarrow +\infty$. Thus $R$ is
uniquely determined for any value of $\kappa\rho_m/\beta\equiv x$
through Eq.(\ref{contract}). Let us denote it simply as
$R=R(\kappa\rho_m/\beta)=R(x)$. Note that irrespective of the
precise form of the relation $R(x)$, this is just an algebraic
relation. Thus for a given $T$, there is no instabilities present
in the metric formulation of the $1/R$ theory indicated by Dolgov
and Kawasaki \cite{Dolgov}, whose origin is due to the fact that
$R$ is determined by a differential equation for a given $T$. To
simply discussion, we will assume $\beta>0$ from now on. Note that
when $\alpha=\beta$, the vacuum solution $R_0\equiv R(0)$ can be
solved exactly as $R_0=-\alpha$.

From the conservation equation $\dot{\rho_m}+3H\rho_m=0$ we can
get
\begin{equation}
\partial_0L'=\frac{3}{(R(x)/\beta)^2-2R(x)/\beta}(\frac{\kappa\rho_m}{\beta})H\equiv F(x)H\label{L'}
\end{equation}
Substituting this to Eq.(\ref{aa}) we can get the Modified
Friedmann equation:
\begin{equation}
H^2=\frac{\kappa\rho_m+2\kappa\rho_r-\beta(\frac{R}{\beta}-\ln\frac{R}{-\alpha})}
{(1-\frac{\beta}{R})(6+3F(x)(1+\frac{1}{2}F(x)))} \label{MF}
\end{equation}

It can be seen from equations (\ref{contract}), (\ref{L'}) and
(\ref{MF}) that when $\beta\rightarrow 0$, the MF equation will
reduce continuously to the standard Friedmann equation. Thus, the
$\ln R$ modification is a smooth and continuous modification.

Let us first discuss the cosmological evolution without matter and
radiation. Define the parameter $n$ as $R_0=-\alpha e^{-n}$.
Substitute this to the vacuum field equation $f(R)=0$, we can get
$\alpha=e^n(2n+1)\beta$ and $R_0=-(2n+1)\beta$. Substitute those
to the vacuum MF equation and set $t=0$, we have
\begin{equation}
H_0^2=\frac{\beta(n+1)}{6(1+\frac{1}{2n+1})}\label{}
\end{equation}
Thus when $\beta\sim H_0^2\sim(10^{-33}eV)^2$ and $n>-1/2$, the
$\ln R$ modified gravity can indeed drive a current exponential
acceleration compatible with the observation. The role of the
parameter $\beta$ is similar to a cosmological constant or the
coefficient of the $1/R$ term in the $1/R$ gravity \cite{Wang1}.

When the energy density of dust can not be neglected, i.e.
$\kappa\rho_m/\beta\gg 1$, $F(x)\sim0$ and if $\alpha$ satisfies
$|\ln(\kappa\rho_m/\alpha)|\ll\kappa\rho_m/\beta$, i.e.
$\exp(-\kappa\rho_m/\beta)\ll\alpha/\beta\ll\exp(\kappa\rho_m/\beta)$,
from Eq.(\ref{contract}), $R\sim\ - \kappa\rho_m$. Then the MF
equation (\ref{MF}) reduces to the standard Friedmann equation
\begin{equation}
H^2=\frac{\kappa}{3}(\rho_m+\rho_r)\label{}
\end{equation}
Thus if
$\exp(-\kappa\rho_{m,BBN}/\beta)\ll\alpha/\beta\ll\exp(\kappa\rho_{m,BBN}/\beta)$,
where $\rho_{m,BBN}$ is the energy density of dust in the epoch of
BBN, the $\ln R$ gravity can be consistent with the BBN
constraints on the form of Friedmann equation \cite{Carroll-BBN}.
One possible choice is $\alpha=\beta$, for which the vacuum
solution can be solved exactly $R_0=-\alpha$. Since $\beta\sim
H_0^2$, the condition $\kappa\rho_m/\beta\gg 1$ breaks down only
in recent cosmological time. Thus the universe evolves in the
standard way until recently, when $\ln R$ term begins to dominate
and drives the observed cosmic acceleration.

\textbf{3. Scalar-tensor formulation of the model}

\begin{figure}
  \includegraphics[width=0.8\columnwidth]{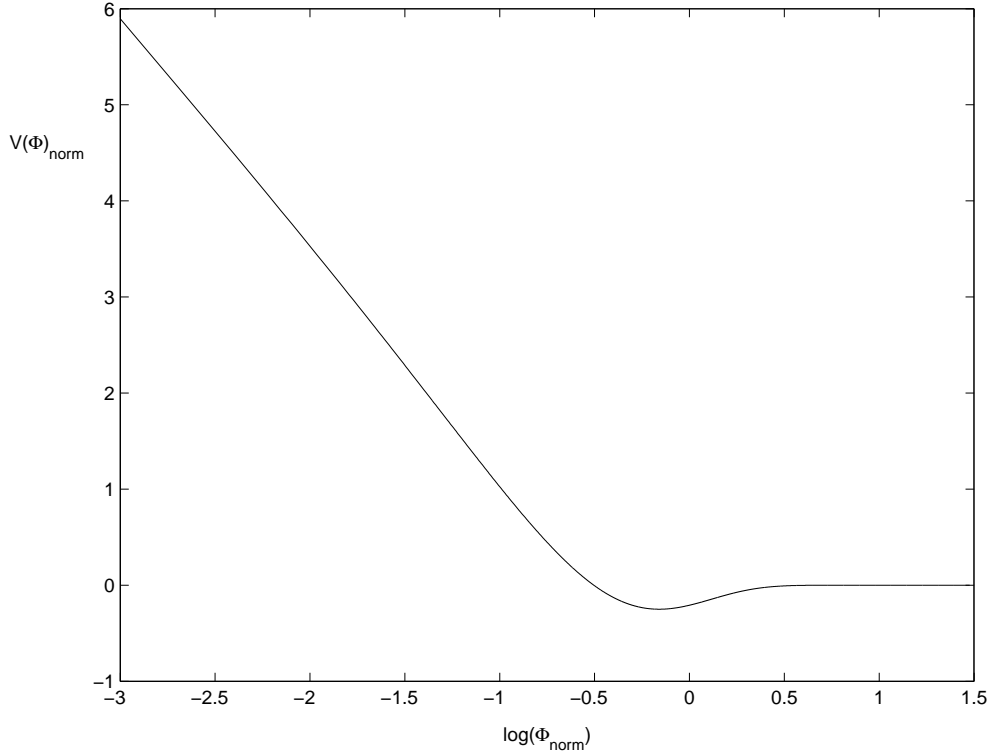}
  \caption{The scalar potential given by Eq.(\ref{poten}) for $\alpha=\beta$. $\Phi_{norm}\equiv\sqrt{2\kappa/3}\Phi$ and $V_{norm}\equiv
  (2\kappa/\beta)V$.}\label{1}
\end{figure}

\begin{figure}
  \includegraphics[width=0.8\columnwidth]{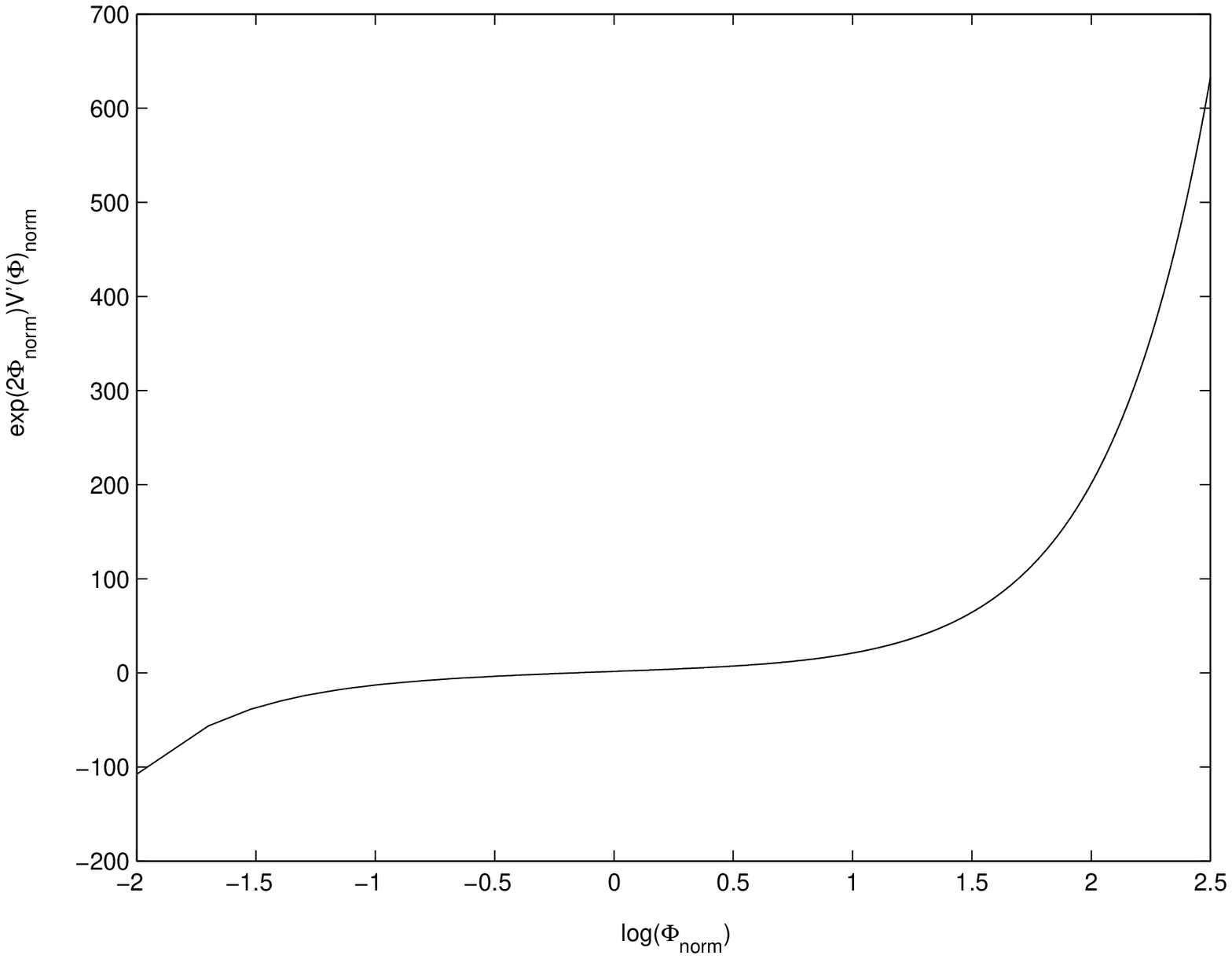}
  \caption{Derivative of the potential for $\alpha=\beta$, from which combined with the field equation (\ref{3.7}) we can
  determine the evolution of $\Phi$. $\Phi_{norm}\equiv\sqrt{2\kappa/3}\Phi$ and $V_{norm}\equiv
  (2\kappa/\beta)V$.}\label{1}
\end{figure}

Recently, Flanagan \cite{Flanagan} derived the equivalent
scalar-tensor theory of the Palatini form of modified gravity
theory. We adopt his formalism and apply it to the $\ln R$ theory.

Following Flanagan, the $\ln R$ theory is equivalent to the
theory:
\begin{equation}
\tilde{S}[\tilde{g}_{\mu\nu}, \Phi, \psi_m]=\int
d^4x\sqrt{-\tilde{g}}[-\frac{\tilde{R}}{2\kappa}-V(\Phi)]+S_m[\exp(-\sqrt{\frac{2\kappa}{3}}\Phi)\tilde{g}_{\mu\nu},
\psi_m]\label{}
\end{equation}
where
$\tilde{g}_{\mu\nu}=\exp(\sqrt{\frac{2\kappa}{3}}\Phi)g_{\mu\nu}$
is the metric in Einstein-frame \cite{Magnano}, $\tilde{R}$ is the
scalar curvature associated with $\tilde{g}_{\mu\nu}$, $\psi_m$ is
the matter field and $\Phi$ is a fictitious scalar field that can
be deleted from the field equations.

The potential $V$ can be obtained by the standard procedure
\cite{Flanagan, Odintsov2}
\begin{equation}
V(\Phi)=\frac{\beta}{2\kappa}[-1+\ln\frac{\beta}{\alpha}-\ln(\exp(\sqrt{\frac{2\kappa}{3}}\Phi)-1)]
\exp(-2\sqrt{\frac{2\kappa}{3}}\Phi)\label{poten}
\end{equation}
See Fig.1 for the case of $\alpha=\beta$. Since the $\alpha$
appears in the expression of $V$ only as the constant term $\ln
(\beta/\alpha)$, other cases would not differ from it essentially.

The field equations are
\begin{equation}
\tilde{G}_{\mu\nu}=-\kappa[V(\Phi)\tilde{g}_{\mu\nu}+\exp(\sqrt{\frac{2\kappa}{3}}\Phi)T_{\mu\nu}]\label{3.6}
\end{equation}
and
\begin{equation}
V'(\Phi)=-\sqrt{\frac{\kappa}{6}}\exp(-2\sqrt{\frac{2\kappa}{3}}\Phi)T=
\sqrt{\frac{\kappa}{6}}\exp(-2\sqrt{\frac{2\kappa}{3}}\Phi)\rho_m\label{3.7}
\end{equation}
where $T_{\mu\nu}$ is the Jordan-frame energy-momentum tensor
defined by Eq.(\ref{2.3}) and $T=g^{\mu\nu}T_{\mu\nu}$.

We can read off the evolution of $\Phi$ from Eq.(\ref{3.7}) (see
Fig.2). In early universe, when $\rho_m$ is large, $\Phi$ locates
at large value; then as the universe evolves, while $\rho_m$
dilutes to smaller and smaller value, correspondingly, $\Phi$
rolls down to the absolute minimum point of the potential at
roughly $\sqrt{2\kappa/3}\Phi\sim 0.7$, at which it can drive an
exponential acceleration expansion.

From Fig.1, we can see that the energy scale of the absolute
minimum of $V$ is of order $\beta/\kappa$ and as shown in Sec.2,
$\beta\sim (10^{-33}eV)^2$. Thus if we assume that the $\ln R$
theory is applicable in small scales such as the electron-electron
scattering scale, there will be a severe conflict with particle
experiment as shown explicitly by Flanagan \cite{Flanagan} for the
$1/R$ gravity. However, those modified gravity theory can not be
fundamental. They are effective theories. If it can be shown that
their cut-off scale is much larger than the electron-electron
scattering scale, the conflict will be fixed. This stimulates us
to pursue their origin from more fundamental theory (see
Ref.\cite{Odintsov3} for such an effort for the $1/R$ gravity). A
large cut-off scale (or a small cut-off energy) is possible for
modified gravity, e.g. for the effective field theory of massive
gravity, Nima Arkani-Hamed et al. \cite{Arkani-Hamed} showed that
the cut-off energy is $(m_g^4M_{Pl})^{1/5}$, where $m_g$ is the
mass of the graviton. This is much lower than the Plank scale, and
correspondingly, its cut-off length scale is much larger than the
Plank length.

In Ref.\cite{Odintsov}, Nojiri and Odintsov also suggested an
extension of the $\ln R$ theory, for which the modified
Einstein-Hilbert action reads as:
\begin{equation}
L(R)=R-\beta\ln \frac{R}{-\alpha}+\gamma R^m\label{3.8}
\end{equation}
We would not discuss this model in detail in this paper. We just
note one thing about it. It would correspond to an unique
equivalent scalar-tensor formulation if the following equation for
$\phi$ has an unique solution for any value of $\Phi$, see
Ref.\cite{Flanagan, Odintsov2}:
\begin{equation}
m\gamma\phi^m-(\exp(\sqrt{\frac{2\kappa}{3}}\Phi)-1)\phi-\beta=0\label{}
\end{equation}
Obviously for $m>1$, this is generally not the case. Thus,
generally, the model (\ref{3.8}) would not have a well-defined
equivalent scalar-tensor theory. What does this imply? According
to the analysis of Magnano and Sokolowski \cite{Magnano}, this is
a strong indication that the original theory is unphysical. Also,
for the $R+1/R+R^2$ theory, for which the same phenomena appears,
\'{E}anna Flanagan \cite{private2} showed that this may imply that
the theory has not a well-behaved initial-value formulation. But
as indicated by Sergei Odintsov \cite{private}, this maybe not
completely the case. The reason is that it is still unclear which
of Einstein or Jordan frame is physical one. For instance, on the
classical level the results obtained in these frames (when
transformation to equivalent theory exists) are identical even for
braneworlds \cite{Odintsov4}. Of course, on quantum level it is
well-known (see explicit examples for quantum dilatonic gravity
\cite{Odintsov5, Buchbinder, Grumiller}) that even classically
equivalent theories are not equivalent on quantum level. Hence,
the fact that metric theory does not have equivalent classical
representation as scalar- tensor theory does not mean that it is
ruled out as physical theory.

\textbf{4. Conclusions and Discussions}

In this paper we discussed the Palatini formation of the modified
gravity with a $\ln R$ term suggested by Nojiri and Odintsov
\cite{Odintsov}. We showed that in the Palatini form, the $\ln R$
gravity can drive a current exponential accelerated expansion and
it reduces to the standard Friedmann evolution for high redshift
region. We discussed the equivalent scalar-tensor formation. We
indicated that the $\ln R$ gravity may still have a conflict with
electron-electron scattering experiment which stimulates us to
pursue a more fundamental theory which can give the $\ln R$
gravity as an effective theory. Finally, we discussed a problem
faced with the extension of the $\ln R$ gravity by adding $R^m$
terms. It is clear that many works still need to be done to see
whether the idea of modifying gravity to achieve cosmic
acceleration in stead of dark energy is viable.

On gravity theory itself, especially the reasonable form of a
quantum gravity is also challenging. With many discussions for
extended gravity models \cite{Odintsov, Wang1, Wang2, Carroll,
Lue, Dvali}, we expect the two tales originate from one truth
 to be discovered.

\textbf{Acknowledgements}

We would especially like to thank Sergei Odintsov for stimulating
this work and many helpful suggestions. We would also like to
thank \'{E}anna Flanagan, Nadeem Haque, Shin'ichi Nojiri for
helpful discussions. This work is partly supported by China NSF,
Doctoral Foundation of National Education Ministry and ICSC-World
lab. scholarship.

\end{document}